\documentstyle[12pt]{book}
\begin{document}

\sloppy
\raggedbottom

\chapter* 
{AN INVITATION TO ALGORITHMIC INFORMATION THEORY\footnote
{\it Lecture given Wednesday 24 April 1996 at a Computer Science
Colloquium at the University of New Mexico.  The lecture was
videotaped; this is an edited transcript.}}
\markright
{An Invitation to Algorithmic Information Theory}
\addcontentsline{toc}{chapter}
{An invitation to algorithmic information theory}

\section*{}G. J. Chaitin\\IBM Research, P.O. Box 218, Yorktown Heights, NY
10598\\chaitin@watson.ibm.com\\http://www.research.ibm.com/people/c/chaitin

\section*{Abstract}

I'll outline the latest version of my limits of math course.
The purpose of this course is to illustrate the proofs of
the key information-theoretic incompleteness theorems of
algorithmic information theory by means of algorithms
written in a specially designed version of LISP\@.
The course is now written in HTML with Java applets, and
is available at http://www.research.ibm.com/people/c/chaitin/lm.
The LISP now used is much friendlier than before, and because
its interpreter is a Java applet it will run in the Netscape
browser as you browse my limits of math Web site.

\section*{Introduction}

Hi everybody!  It's a great pleasure being back in this beautiful,
beautiful state.  You guys were nice enough to invite me up a bunch of
times in the past.  And I've usually tried to explain general ideas.
I thought this time I'd try to give a different kind of a talk.

I've been working for several years on a project, on a course I call
``The Limits of Mathematics.''  And this course in the past three
years has gone through a number of different versions.  It has a lot
of software attached to it.  And I've been changing this software.
I've been trying to explain this software to people as the course.  In
fact a few weeks from now I'll be giving this course in Rovaniemi,
Finland, on the arctic circle, for two weeks.

The way this course is now is that it's on this Web site that I gave
you guys the URL for in the abstract.  It's an HTML document and it
has a lot of LISP code hanging off of it, and just to make life worse
this is not normal LISP\@.  It's pretty close to pure LISP, you know,
the heart of LISP that everybody learns.  But I've had to add a few
features to LISP\@.  So this stuff is all available and you're welcome
to play with it.  The course is there.

What's also there by the way on my Web site are the transcripts
of the previous talks I've given here at the University of New Mexico.
One of them is ``The Berry Paradox,'' it's in HTML.  Another one is
called ``Randomness in Arithmetic and the Decline and Fall of
Reductionism in Pure Mathematics.''  That's another talk that I
gave here.  And there's a talk that I should have given here but it
didn't go well.  The version I gave at the Santa Fe Institute is
there.  It's called ``How to Run Algorithmic Information Theory on a
Computer.''  So these are three very understandable papers and I give
them out when I give this course.

Now there is a small question: Once a paper is published in a
journal are you allowed to have it anymore on your Web site?  At my
lab at this moment the view is that you should erase it.  Los Alamos
has a wonderful physics preprint server and they don't erase anything.
And all over the academic community everybody has all their key papers
on their Web site even after they're published, right?  So I
don't know.  At this moment all this stuff is on my Web site.  Some of
it may disappear temporarily while the matter is discussed.

\section*{Recursive function theory revisited}

My work basically is doing recursive function theory at the kind of
level it was done in the early days in the 1930's before it got
technical.  At the kind of level that G\"odel and Turing were doing it
where the concern was what are the limits of mathematical reasoning.
And basically I've added two new things.  This is sort of revisited
sixty years later, this stuff.  I've added two new things.

One new thing I add to the stew is that an idea that was missing
which is significant is the idea of program-size complexity, or how
big a program is, how many bits of information there are.  Not the
time!  In that sense it's just like the old recursive function theory
and Turing machines.  I don't care about time.  But I do care about
the size of a program in bits.  Okay?  So that's one new idea.

The other new idea I've added just in the past three years is that I
want to really be able to run the programs on interesting examples on
a computer.  In other words, it's not enough to wave your hands and
talk about an algorithm.  I not only want a universal Turing machine
that I can prove theorems with, but one that is fun to actually
program and that you can run interesting examples on with a computer.

John McCarthy invented LISP not just for use in AI [artificial
intelligence], not just for use as a practical language.  If you look
at his 1960 paper on LISP in the {\it Communications of the ACM,} he
said, ``This is a better universal Turing machine.  Let's do recursive
function theory that way!''\footnote{Of course these aren't
McCarthy's words, it's my paraphrase.} And the funny thing is that
nobody except me has really I think taken that seriously.  And the
reason of course is that theoreticians didn't really much care about
programming, about really playing with the computer.  So the thing
I've added is I think nowadays you have no excuse!  If a theory has to
do with the size of computer programs, you want to damn well see the
programs, you want to be able to run them, it ought to be a
programming language that is easy to use.

So I've done that using LISP because LISP is simple enough, LISP is in
the intersection between theoretical and practical programming.
Lambda calculus is even simpler and more elegant than LISP, but it's
unusable.  Pure lambda calculus with combinators $A$ and $K$,
it's beautifully elegant, but you can't really run programs that way,
they're too slow.  So I would say that I believe that LISP is a
powerful language that can actually be used on a computer to run stuff
in a finite amount of time, but on the other hand it's simple enough
and elegant enough that you can prove theorems about it.  So that's
the kind of game that I've been playing.  Okay, so you'll find this on
my Web site.

And as general background then I suggest these three papers: ``The
Berry Paradox,'' which was published in the first issue of {\it
Complexity\/} actually.  This other lecture on ``The Decline and Fall
of Reductionism...'' which was published, among other places, by
Oxford University Press in {\it Nature's Imagination,} that's the name
of the volume.  Oh there's also a paper called ``A New Version of
Algorithmic Information Theory,'' which is more technical, which was
published in the latest issue, the current issue, the fourth issue of
{\it Complexity\/} magazine, the one which just came out.  And the
paper ``How to Run Algorithmic Information Theory on a Computer''
which kicks around these ideas is going to come out in a future issue
of {\it Complexity.} But for now they're all on the Web site, until
they get thrown out!

\section*{My new LISP \& complexity measure}

Okay, so let me leap in and give you the general idea.  What is this
big new thing you add to LISP, this change you make to LISP to be able
to talk about program-size complexity?  Well, you could just take LISP
expressions and measure how big are they in characters, right?  That's
not a bad complexity measure.  But it's not the best complexity
measure.

The best complexity measure says, you have a LISP S-expression, which
is a computer program, and then you have binary data on the side:
\begin{verbatim}
   lisp       binary
   S-exp      data
\end{verbatim}
The problem with measuring the size of a LISP program is that due to
LISP syntax the LISP S-expression is redundant.  So you really want to
add to the LISP S-expression, which is a powerful way to express an
algorithm, some raw binary data.  That's like having data on a tape in
addition to your program.  So in the data the bits can be 0 and 1
independently, and there's some way that the program can get access to
the data, and then you look at the size of the whole thing.

So basically the universal Turing machine I'm using as my yardstick to
talk about program size, to use to measure program size, has programs
which are big binary programs:
\begin{verbatim}
   BINARY PROGRAM
   lisp       binary
   S-exp      data
\end{verbatim}
That sounds bad, right?  We have this awful machine language in
binary.  You're not going to want to write in binary.  Well, it's not
so bad.  The beginning of a program is a LISP S-expression converted
to binary, 8 bits per character.  And you know where the LISP
S-expression ends, because by convention there is always a special
ASCII character at the end.
\begin{verbatim}
   BINARY PROGRAM
   lisp       binary
   S-exp      data
  8 bits/char
\end{verbatim}

So the universal Turing machine starts reading the binary tape bit by
bit.  It doesn't sort of gobble it all in and start running.  It reads
each bit of the program as it needs to---that's very important.  And
it starts off by reading a LISP S-expression, and then every time the
LISP S-expression says ``Read another bit,'' it gets it from what's
left on the tape, that is, in the binary program.  And if the
S-expression attempts to read beyond the end of the binary data, the
program is a failure and is aborted.  So that's how it goes.  And then
you just add the total number of bits, which is 8 times the number of
characters in the S-expression plus the number of bits in the data.
That gives you the total number of bits in the program, okay?

So that's done because just using the size in characters of a LISP
S-expression (or multiplying it by some factor to have it in bits) is
no good.  LISP syntax is pretty simple, but even so there's redundancy
there.  So you've got to add this extra binary data on the side.

\section*{Programming my UTM in LISP}

So let me show you my universal Turing machine programmed out in my
new LISP\@.  When I came here a year ago, I had a LISP, it was pretty
awful.  It was a LISP where atoms and variable names were only one
character long.  And it was a LISP where there was no arithmetic.  The
way you had to do integers was as lists of 0's and 1's and you had to
define, you had to program out arithmetic.  I thought this was cute,
this is a fun game to play once in your life, but the problem is it's
an obstacle if I want to explain this to anybody else!  So now I have
a LISP like all other LISP's with long names for atoms, and my
universal Turing machine is now programmed like this.

Define $U$ of $p$, $U$ is the function, $p$ is the program, as
follows:
\begin{verbatim}
   define (U p)
   cadr try no-time-limit
            'eval read-exp
            p
\end{verbatim}
Now this is it, this is my universal Turing machine that I described
in words.  {\tt Try} is a primitive function in this LISP sort of like
{\tt eval} in normal LISP\@.  {\tt Try} has three arguments, here they
are:
\begin{verbatim}
            no-time-limit
            'eval read-exp
            p
\end{verbatim}
So this, argument 2,
\begin{verbatim}
            'eval read-exp
\end{verbatim}
is a LISP expression that we're going to evaluate with a time limit.
In fact there's no time limit here:
\begin{verbatim}
            no-time-limit
\end{verbatim}
You also have associated binary data.  {\tt Try} is the way that you
give a LISP expression binary data.  So this
\begin{verbatim}
            p
\end{verbatim}
is the binary data, it's the program for the universal Turing machine.
It's a list of 0's and 1's.  This is {\tt try}'s third argument.

Oh by the way, in normal LISP you have to put lots of parentheses.  In
my LISP I leave out a lot of them because in my LISP every built-in
function has a fixed number of arguments, so I don't bother to write
all the parentheses.

Back to the LISP code for my UTM.
\begin{verbatim}
            (read-exp)
\end{verbatim}
is a primitive function with no arguments.  So we're trying to
evaluate this
\begin{verbatim}
            ('(eval(read-exp)))
\end{verbatim}
normally with a time limit, but in this case with no time limit.  And
{\tt read-exp} reads a LISP expression from the beginning of the
binary data {\tt p}.  And then we evaluate it, in other words, we run
it.  And while you're running it, if within the LISP expression that
you read you ask for more binary data, you get it from what's left of
the binary data.

It works!  {\tt Try} does exactly what it should, I added {\tt try} to
normal LISP just for this.  And {\tt try} is the real difference
between normal LISP and my LISP\@.  Okay, so this
\begin{verbatim}
  (define (U p)
  (cadr (try no-time-limit
             ('(eval(read-exp)))
             p)))
\end{verbatim}
is my universal Turing machine, which I think is not a very big,
complicated program.

And the nice thing about this is that you see lots of examples in my
course.  You have there a Java applet for the LISP interpreter.  Which
means you can write LISP and see what it does.  But most of the course
is ``canned'' LISP\@.  What I do is I define some big LISP functions
then I run them on examples.  And there are lots of comments saying
why I'm doing it and what it proves.

Actually there are two versions of each of these canned LISP runs, one
of them with no comments and running it just on the case that counts,
and another one with all the comments in the world and with lots of
test cases for all the auxiliary functions.  I do this because without
comments a program is incomprehensible, but when you put all the
comments in it becomes incomprehensible for another reason---it's so
big that you can't find anything.  So I present it both ways.

Okay, so one of the things you can do is you can actually get a
program and run it on this machine.  So how do you do it?  Well, you
take a LISP expression, and then you want to convert it to binary, so
I have a primitive function for doing that, and then you concatenate
it to the binary data that it needs, and then you feed this into the
function $U$ of $p$ that we defined.  And in my Web site I have runs
of the universal Turing machine that show how it works.

\section*{A simple program for U that proves a theorem}

Let me show you a particularly interesting example of this, and I'm
also going to prove a theorem, a very simple theorem about my version
of program-size complexity.  This theorem says that the program-size
complexity of a pair of objects is bounded by the sum of the
individual program-size complexities of the two objects plus a
constant:
\[
   H(x,y) \le H(x) + H(y) + c .
\]

What this means is this.  $H(x)$ is the size in bits of the smallest
program for the universal machine $U$ of $p$ that calculates $x$; $x$
is an S-expression.  And $H(y)$ is the size of the one that calculates
$y$.  And putting them together and adding a fixed number of bits you
get a program that calculates the pair $(x,y)$, which won't have a
comma by the way, since we're doing LISP, which is without commas.

So how does this work?  Well, let me show you what the program looks
like.  What you do is you take the minimum size program for $x$, the
minimum size program for $y$, and I'll call that $x^\ast$ and
$y^\ast$, and you concatenate in front a prefix $\phi_c$, a magic prefix
that's going to be $c$ bits long, and I'll show you this prefix.  This
gives you
\[
   \phi_c \, x^\ast y^\ast .
\]
Then when you give it to the universal machine $U$ of $p$ it's going
to produce the pair $(x,y)$.

Now what is this prefix $\phi_c$ that you're concatenating in front of
these two bit strings $x^\ast$ and $y^\ast$?  Actually you can put any
program for $x$ and any program for $y$ here in $\phi_c \, x^\ast y^\ast$.
I don't know the minimum ones.  You'll see that this prefix works
because if you give it any program $x^\ast$ to calculate $x$ and any
program $y^\ast$ to calculate $y$, this prefix $\phi_c$ when you feed it
to my universal machine $U$ of $p$ is going to end up being a program
that calculates the pair $(x,y)$.

So what is this prefix $\phi_c$?  Well, the prefix is this.  Start with
\begin{verbatim}
  (cons (eval (read-exp))
  (cons (eval (read-exp))
        nil))
\end{verbatim}
Here I'm putting in all the parentheses.  Then convert this to a bit
string $\phi_c$.  So you write this expression and use a primitive
function to convert it to a bit string and then you use {\tt append}
to concatenate these lists of bits and get this long bit string:
\[
   \phi_c \, x^\ast y^\ast .
\]
And then you feed it into $U$ and it produces $(x,y)$.

How does this
\begin{verbatim}
  (cons (eval (read-exp))
  (cons (eval (read-exp))
        nil))
\end{verbatim}
work?  It says, ``Read from the binary data.''  The binary data when
$\phi_c$ is running is going to be $x^\ast y^\ast$, the rest of the
program.  So it's going to read and run the program $x^\ast$ to
calculate $x$, and that's going to give it $x$.  And then it's going
to read off of the rest of the binary data and run the program
$y^\ast$ to calculate $y$, and that's going to give it $y$.  And then
it's going to {\tt cons} these two things $x$ and $y$ up into a pair.
Okay?

So it's the number of bits in this S-expression $\phi_c$
\begin{verbatim}
  (cons (eval (read-exp))
  (cons (eval (read-exp))
        nil))
\end{verbatim}
which is the constant $c$ in
\[
   H(x,y) \le H(x) + H(y) + c .
\]
So count the number of characters here in $\phi_c$,
multiply by 8, and that's the constant $c$ in this theorem.
Okay?  So this is pretty straight-forward.

\section*{``Computing'' the halting probability $\Omega$}

Now, what do I do next in my course on my Web site?  Okay, so I've
defined this universal Turing machine $U$ of $p$.  The next major step
in my theory is to define the halting probability $\Omega$.  This is a
number that a lot of people get excited about [laughter], the halting
probability.  Well, I'm going to write down a program to calculate the
halting probability now.  Let me actually write it down for you!  Here
it goes.  Let's write this program out.  So I'm going to write a LISP
program as we go.

Okay, so we're going to define a function that counts how many
programs halt, let's call it {\tt count-halt}, which has two
arguments, one is a {\tt prefix}, and the other is {\tt bits-left}.
\begin{verbatim}
  define (count-halt prefix bits-left)
\end{verbatim}
What's the idea?  I'm going to count how many program for my universal
Turing machine halt... Oh, I should put in time... Let me put
in {\tt time}, there's another argument here:
\begin{verbatim}
  define (count-halt time prefix bits-left)
\end{verbatim}
So I'm going to count how many programs halt on my universal Turing
machine that have this {\tt prefix} with {\tt bits-left} bits added on
afterwards, within this amount of time.  So this {\tt count-halt} is
the auxiliary function that I use.  Okay?

So the first thing to show you is how you get from this the halting
probability.  So before I define the auxiliary function, let me show
you how to define a function {\tt omega}, the main function, that uses
this auxiliary function.  It's only one main function with one
auxiliary function.  Let's call it {\tt omega} of $n$:
\begin{verbatim}
  define (omega n)
\end{verbatim}
This is the $n$th approximation, the $n$th lower bound on the halting
probability.  As $n$ goes to infinity, this will give you the halting
probability in the limit from below.  The problem is it converges
very, very, very, very slowly!  [Laughter] Noncomputably slowly!
Thanks for the laughs.

Let's define this function.  Well, what I want to do is I want to {\tt
cons} up a rational number.  And the way I do it is, I count how many
programs halt within time $n$ that have as prefix {\tt nil}, the empty
bit string, and that add $n$ bits to it.
\begin{verbatim}
  define (omega n)
  cons (count-halt n nil n)
\end{verbatim}
And I {\tt cons} that with division.
\begin{verbatim}
  define (omega n)
  cons (count-halt n nil n)
  cons /
\end{verbatim}
And {\tt cons} that with, I use \verb|^| for power, 2 raised to the
power $n$, and {\tt nil}.
\begin{verbatim}
  (define (omega n)
  (cons (count-halt n nil n)
  (cons /
  (cons (^ 2 n)
        nil))))
\end{verbatim}

This depends on the representation you pick for rational numbers; I
just write out a fraction.  And so {\tt (omega n)} is a triple, I'm
creating a triple.  So it's (the number of programs $n$ bits in size
that halt within time $n$) divided by $2^n$.  Okay?

So that's the main function.  Now let me define the auxiliary
function.  The auxiliary function goes like this.  You start off with
the empty prefix {\tt nil} and recursively you start adding bits until
you get a full $n$-bit program and then you run it for time $n$ and
you see whether it halts.  Now how do you program that out in LISP?
Well, recursively it's very easy, you go like this.

Let's first see if {\tt bits-left} is not equal to 0.  If {\tt
bits-left} is greater than 0, then add a 0 bit to the prefix and
subtract one from the number of bits left, see how many of these
programs halt within time $n$, and add that to the same thing when you
{\tt append} a 1 to the prefix instead of a 0.

So we've got this:
\begin{verbatim}
  define (count-halt time prefix bits-left)
  if > bits-left 0
  + (count-halt time (append prefix '(0)) (- bits-left 1))
    (count-halt time (append prefix '(1)) (- bits-left 1))
\end{verbatim}

I'll put parentheses in occasionally and other times not.  [Laughter]
Well, actually in my LISP you don't put them in for primitive
functions and the interpreter adds them, and writes it out for you
just so that you can check that it's what you meant.

So this is the recursion.  This is going to make {\tt count-halt} look
at all $n$-bit programs.  And what happens when it finally has no bits
left to add?  It started off with $n$ bits to append to the prefix,
and with the empty prefix.  So now you've finally got an $n$-bit
string, you get all possible $n$-bit strings.  So then you take the
$n$-bit string, and you {\tt try} for time $n$ {\tt eval read-exp}
applied to the $n$-bit data string which is the prefix.  If this {\tt
try} is a {\tt success} then {\tt count-halt} is a 1 otherwise it's a
0.

So now we've got this:
\begin{verbatim}
  define (count-halt time prefix bits-left)
  if > bits-left 0
  + (count-halt time (append prefix '(0)) (- bits-left 1))
    (count-halt time (append prefix '(1)) (- bits-left 1))
  if = success car (try time
                        'eval read-exp
                        prefix)
     1
     0
\end{verbatim}

This recursion is going out and seeing whether all possible $n$-bit
programs halt or not.  When you've got the whole $n$-bit program, how
do you see whether it halts or not?  You {\tt try} running it for the
time that you were given, which is going to be $n$ for {\tt omega} of
$n$.  And this part
\begin{verbatim}
                   (try time
                        'eval read-exp
                        prefix)
\end{verbatim}
does that.  It's taking the prefix as the S-expression and the binary
data of a program for my universal Turing machine.  It's reading an
S-expression off of the beginning of the prefix and it's running the
S-expression with the rest of the prefix as binary data.  Okay?

I don't really have the time to explain this properly.  If you were
giving this to students in a class you'd probably take a whole class
to explain this.  And then the students would go and they would play
with it and they would use it.  And if you look at my Web site you'll
find that I'm running {\tt count-halt} on examples to show that it
works.  I also have a different definition of {\tt omega} of $n$ which
is more traditional, which builds up the list of all $n$-bit strings,
and then sees which halt, and then counts how many halt.  But I think
that what I have here is more elegant, it's really all you need.

Let me say it in words again.

What you see here in this version of LISP is the $n$th lower bound on
$\Omega$.  With increasing positive integers $n$, {\tt omega} of $n$
gives you better and better lower bounds on the halting probability.
The problem is that you never know how far out to go to get the
halting probability with a given degree of accuracy.  To get $n$ bits
of the halting probability right, you have to go out Busy Beaver
function of $n$, that is basically the idea.

You look at all $n$-bit programs, as prefix you start off with the
empty list of bits, and the bits left to add is $n$ initially, and
you're going to count how many of them halt within time $n$.  That's
what
\begin{verbatim}
       (count-halt n nil n)
\end{verbatim}
does.

\section*{Discussion of the program for $\Omega$}

Let's look at this conceptually and let's look at it pedagogically.
Pedagogically, I certainly learned a lot writing this program, right?
I had to invent the programming language, I had to write the
interpreter!  So I learned a lot from this.  Is this a good way to
transmit an idea to a student?  I don't know!  Reading somebody else's
code is not great, right?

You really understand an algorithm when you've programmed it and tried
it on examples and debugged it yourself.  So I think that to use this
in a class effectively you need exercises.  You need to ask the
students to do variations on what you've presented, so they make it
their own.  I don't do this here.  This is a very concise version of
the course.

If someone really wanted to use this as a course, they could start
with this, which it is at least feasible that someone could
understand.  Before it wasn't.  I could understand it.  A few very
committed people could sort of try to think they understood it, but I
think there's more of a chance with this, it looks more like normal
LISP\@.  So that's from the pedagogic point of view.

Now let's look at this philosophically.  The point is this.  This
program is $\Omega$'s definition.  This is really it.  So if this
number is going to be very mind-boggling, you'd like to pin down very
concretely how it's defined.  And we have.

You'd like to pin down very concretely how $\Omega$'s defined before
you start being astonished by it.  Here $\Omega$ is really very
concrete and you can actually run {\tt (omega 0)}, {\tt (omega 1)},
{\tt (omega 2)}, ... Of course the time grows exponentially as $n$
goes up, but maybe you can get to one interesting example.  You can
try variations of this program to convince yourself that you
understand it.  And you can try debugging {\tt count-halt} separately
by running examples to show that it works.  That's how to convince
yourself that this works: run lots of examples.  I haven't tried
giving a formal proof that this program is correct.  I wonder if that
would be an interesting project?

Okay, now I once tried explaining this to an astrophysicist who
amazingly enough was briefly interested [laughter], a very
bright guy.  He was so bright that he could understand the previous
version of the program for $\Omega$, the one in which the names of
built-in functions are a single character, and where you had to
program out arithmetic, you didn't have positive integers like we do
here.  He took the definition of {\tt (omega n)} and {\tt count-halt},
printed them on a small sheet of paper, folded it up and put it in his
wallet.  He said that it was like a mantra, ``I have $\Omega$ in my
pocket!''  But it was $\Omega$ for him only because he was very bright
and he took the trouble to go through this course with me.  I think
that this new version is much easier to understand than the one that
he put in his pocket.

\section*{Proof that $\Omega$ is algorithmically irreducible}

Okay, so what is the next thing that one does in this course on the
limits of mathematics?  I've tried to throw out all the inessential
stuff and just give the main ideas.  And now we'll start to see why
this number $\Omega$ is significant.

Well, the next thing I do in the course is I point out that {\tt
(omega 0)}, {\tt (omega 1)}, {\tt (omega 2)}, ... is a monotone
increasing sequence of rational numbers.  So, this isn't constructive,
but we all know that there is a real number which is the least upper
bound of this sequence.  So {\tt (omega n)} defines in a
nonconstructive way a real number $\Omega$.  It's almost constructive.
It's pretty close to being constructive.  Really constructive would be
if you could calculate each digit of $\Omega$, right?  I'm getting
better and better lower bounds, the only thing I don't have is I don't
have a way to calculate how far out to go to get a given degree of
accuracy.

Okay, so this defines a real number that I call $\Omega$.  Now let's
imagine this number being written in base-two binary, so there's going
to be a 0, a binary point, and then there're going to be some bits,
right?
\[
   \Omega = 0.011101\ldots
\]
And let's think of this bit string as a LISP S-expression, it's going
to be a list of 0's and 1's separated by blanks.  And let's define
$\Omega_N$ to be the first $N$ bits of this, the LISP S-expression for
the first $N$ bits of $\Omega$.  It's a list with $N$ elements.

The theorem you want to show is that $H(\Omega_N)$ is pretty
complicated, that there's a lot of information in those first $N$ bits
of $\Omega$.  In fact it's algorithmically irreducible.  You cannot
compress the first $N$ bits of $\Omega$ into a program substantially
less than $N$ bits in size.  The exact result is that
\[
   H(\Omega_N) > N - 8000 .
\]
This turns out to be the case.

Why is there a lot of information in the first $N$ bits of $\Omega$?
It's not difficult to see why.  If you knew the first $N$ bits of
$\Omega$, it solves the halting problem.  It would enable you, slowly,
but it would enable you to solve the halting problem for all programs
up to $N$ bits in size.  So once you do that, you run all the programs
up to $N$ bits in size that halt, you see what they calculate, and
then you just put what they calculate together into a list.  That
gives you something which cannot be done by a program less than or
equal to $N$ bits in size.

And it takes 8000 bits to explain how to do this.  There's a prefix
$\phi_{8000}$ which is 8000 bits long that if you put it in front of a
program that calculates the first $N$ bits of $\Omega$, this gives you
a program to calculate something whose complexity is greater than $N$.
This shows rather concretely that this inequality
\[
   H(\Omega_N) > N - 8000
\]
has to follow.  Okay?

So I'm giving my universal Turing machine $U$ a program which starts
with an 8000-bit prefix $\phi_{8000}$.  That sounds like a lot, but in
fact it's only a thousand characters of LISP\@.  And then next to it I
concatenate any program $\Omega_N^\ast$ for calculating the first $N$
bits of $\Omega$, if somehow I had one.  I don't know where to get
one!  So this is what we've got:
\[
   U( \phi_{8000} \, \Omega_N^\ast ) .
\]
And then you run this and what the machine $U$ does because of
$\phi_{8000}$ is this.  First it reads in and runs $\Omega_N^\ast$ to
calculate the first $N$ bits of $\Omega$.  Then it uses {\tt (omega
n)} to get better and better lower bounds on $\Omega$ until $\Omega_N$
is correct, until it gets the first $N$ bits of $\Omega$ right.  God,
or an oracle, is giving us the program for the first $N$ bits of
$\Omega$!

Once $U$ has calculated {\tt (omega n)} for $n$ high enough that the
first $N$ bits of $\Omega$ are correct, at that point it's seen all
$N$-bit programs that are ever going to halt.  In fact, if they ever
halt, they halt within time $n$.

And then $U$ sees what these programs produce, and it forms the list
of the output of all programs up to $N$ bits in size that halt.  And
to finish up everything, $U$ outputs this list and halts.  In other
words, this list is the value of
\[
   U( \phi_{8000} \, \Omega_N^\ast ) .
\]
It's precisely the list of all LISP S-expressions $x$ with
program-size complexity $H(x) \le N$.

This list cannot itself have program-size complexity $\le N$, because
it can't be contained within itself.  So it can't be the output of a
program less than or equal to $N$ bits in size, it's got to have
program-size complexity greater than $N$.  Therefore this program
$\phi_{8000} \, \Omega_N^\ast$ for producing the list
\[
   U( \phi_{8000} \, \Omega_N^\ast )
\]
must be greater than $N$ bits in size.  In other words, (the size of
any program that calculates the first $N$ bits of $\Omega$) plus 8000
has got to be greater than $N$.  So you get this inequality
\[
   H(\Omega_N) > N - 8000 .
\]
Okay?

And in my Web site I actually show you this program.  I program out in
LISP the algorithm that I just described in words.  You program it
out, it's not a big deal, $\phi_{8000}$ is about one page of LISP, of
my new LISP\@.  But if you put in a lot of comments and run lots of
examples then it's more than one page.  Okay?

Martin Gardner had an explanation of this algorithm in an article in
{\it Scientific American\/} on $\Omega$ that you'll find in one of the
Martin Gardner collections.  In my Web site this algorithm is actually
written out in LISP and you can run it on some examples.  You can
certainly run the auxiliary functions and other pieces of the
algorithm and convince yourself that it works.  But when you put it
all together, you're not usually going to be able to run it on the
real data $\Omega_N$, because the whole point of this is to show that
the bits of $\Omega$ are hard to know, they're algorithmically
irreducible.

By the way, it's easy to see that
\[
   H(\Omega_N) > N - 8000
\]
implies that $\Omega$ is violently non-computable.  Let's compare the
two real numbers $\Omega$ and $\pi$.  $\pi$ has the property that the
string of its first $N$ bits has very small program-size complexity.
Given $N$, you can calculate the first $N$ bits of $\pi$.  So the
first $N$ bits of $\pi$ only have about $\log_2 N$ bits of complexity.
But the first $N$ bits of $\Omega$ are irreducible, can be reduced at
most 7999 bits, maybe down to $N-7999$, we've proved that.

By the way, program-size irreducibility implies statistical
randomness.  So $\Omega$'s irreducibility implies that $\Omega$ is a
normal real number.  In any base, its digits all have the same
limiting relative frequency.

\section*{Proof that you can't even deduce what the bits of $\Omega$
are}

Okay, so the next thing you do with this, and this is the next (and
last) program in this course, is you get an incompleteness result for
$\Omega$ from this.  From this inequality
\[
   H(\Omega_N) > N - 8000
\]
which states that $\Omega$ is algorithmically irreducible.  By the
way, I don't show in the course that this implies statistical
randomness.  That's a detour.  I want to go straight to the
incompleteness result.

(But I think you can sort of wave your hands and try to convince
people that irreducibility implies statistical randomness.  Or you can
actually go and develop algorithmic information theory in excruciating
detail!  But in my course my goal is to get to the major
incompleteness results as quickly as possible, illustrating them with
interesting LISP programs, not to develop the entire theory.  Now you
could also take the proof of every theorem in my Cambridge University
Press book {\it Algorithmic Information Theory\/} and write down a
LISP program for the algorithm in it.  I once started doing that but I
quickly gave it up.  It was too much work, and you don't really want
to see every algorithm in the theory in such detail.  I'm
concentrating on programming in LISP what I think are the fun
algorithms in the theory, especially the ones connected with the basic
incompleteness results, the ones I want to use to present what I think
are the fundamental concepts in my theory.)

So the last program in this course is the one that yields an
incompleteness result for $\Omega$.

Let's start with this question: How do I represent a formal axiomatic
system in LISP?  Well, this is the way I do it.  Think of it as a
black box which every now and then throws out a theorem.  So it's a
program that you start running, and it may or may not halt, normally
it doesn't halt, and every now and then it prints out a theorem.

So why is this a formal axiomatic system?  Well, the idea is that I don't
care about the details of what's going on inside, what the axioms are,
or what the logic used is like.  What I care about is the
criterion, which I think was enunciated rather clearly by Hilbert, that
states that the essence of a formal axiomatic system is that there
should be a proof-checking algorithm.  So if that's the case, you can
run through all possible proofs in size order, see which ones are
correct, and print out all the theorems.  Of course this is
impractical.  The time it takes to do it grows exponentially, and I
wouldn't really want to run such a program.

But you can cheat and you can just have a LISP S-expression which
every now and then outputs a theorem... It won't be the final value...
Every now and then it uses a primitive function, I call it {\tt
display}, but you could also say ``output intermediate result.''  So I
think of a formal axiomatic system as a LISP S-expression that every
now and then calls a particular new LISP primitive function which
outputs an intermediate result.  And that way it puts out a lot of
LISP S-expressions, which are the theorems.

And you can cheat and instead of putting a real formal axiomatic
system in there with a proof-checking algorithm and running through
all possible proofs, which is terribly slow (and never halts),
you can cheat and put in a little example of something that just
{\tt display}'s a few sample theorems and then halts.  That way
you can run a little example and debug algorithms which work with
formal axiomatic systems.

How can algorithms work with these toy formal axiomatic systems?
Let's say that you have a formal axiomatic system.  We've agreed that
it's a LISP program that is going to put out intermediate results
using a new primitive function called {\tt display}.  Actually all
normal LISP's have a function for outputting intermediate results; you
use it for debugging, right?  But it becomes more important in this
framework.  It plays an official role in my computerized version of
algorithmic information theory, because {\tt try} captures all the
intermediate results.  That's very important.

{\tt Try} is a time-bounded execution/evaluation of a LISP expression.
You use {\tt try} to run a program with a time limit, giving the
program raw binary data on the side, and {\bf capturing all the
intermediate output} as well as the final value (if any).  That's very
important.  You see, the new primitive function {\tt try} that I added
to LISP is the way that you can run a formal axiomatic system for a
while and see what are the theorems that it produces.  That's put in
as a primitive function in this LISP\@.  The value returned by {\tt
try} is a 3-element list $( \alpha \, \beta \, \gamma )$.  $\alpha$ is
{\tt success} or {\tt failure}.  $\beta$ is the value of the
expression being tried if the {\tt try} is a success, and is {\tt
out-of-time} or {\tt out-of-data} if the {\tt try} is a failure.  And
$\gamma$ is the list of captured {\tt display}'s, the list of
theorems.  The theorems don't get displayed, they end up in this list
instead.  Okay?

Let me say by the way that this LISP started off as three-hundred
lines of Mathematica.  I invented this LISP using Mathematica as my
programming tool; I wrote the LISP interpreter in Mathematica.  That
way I could play with my LISP and try it out as the design evolved.
Mathematica is the most powerful programming language that I know.
But it's slow.  So the next thing I did was I rewrote the interpreter
for this LISP in C\@.  And it's a thousand lines of C, it's a hundred
times faster than in Mathematica, but the program is incomprehensible
of course.

So the next thing was, I rewrote the interpreter in Java, because by
this time HTML, the World Wide Web, and Java had appeared!  And in
Java it's 750 lines of code.  But I cheat, I didn't take the trouble
to program out {\tt bignum}'s, arbitrarily large integers.  I did that
in C, Mathematica comes with it built in.  And in Java you get
18-digit decimal integers built in, so I said, ``That's enough!''
Probably somebody someday is going to add arbitrarily large integers
to the Java class libraries, it's object oriented, a lot of stuff is
in those libraries.  I don't want to do that work.

So it's 750 lines of Java, and I think that the Java code is pretty
nice.  The C code is incomprehensible!  Like all good C programs, it's
too clever.  You can deal with it while you're writing it, and
immediately afterwards it's already incomprehensible, even for the
programmer!  The Mathematica code is much easier to understand, but
you have to know Mathematica, which is a pretty big language, and you
have to know the subset of it that I'm using, which may not be the
subset that you like to play with normally.

The Java code I have to say I really like.  It's 750 lines of Java.  I
think it's pretty understandable, I think it's pretty clean, and I
think that if I were presenting this as a course I would go through
the Java interpreter with the students.  I think it's important that
they should see the code.  It's only 750 lines of Java, Java's a
fairly reasonable language, and they could make changes to the
interpreter as a way of showing that they understand it.  So it's
there, the Java source code is there on my Web site.

Okay, so what's the next (and the last) program?  The next program
works with a formal axiomatic system, which is just a mechanism for
producing an infinite set of theorems.  It's given to us as a binary
program for $U$, FAS$^\ast$, which is a LISP expression plus binary
data, and we're going to measure its complexity in bits the same way
that we did before.  I'll prove that if the formal axiomatic system
has program-size complexity $N$, then it can enable you to determine,
to prove what is the value of at most $N+15328$ bits of the halting
probability $\Omega$.
\begin{quote}
If a FAS has program-size complexity $N$, then it can enable you to
determine at most $N+15328$ bits of $\Omega$.\footnote
{Perhaps one should refer to this as the {\it logical (or deductive)
irreducibility\/} of $\Omega$, to distinguish it from the {\it
computational (or algorithmic) irreducibility\/} of $\Omega$, namely
the fact that $H(\Omega_N) > N - 8000$.}
\end{quote}

So how do I show this?  Well, I have a Berry paradox kind of proof.
The idea is that if you could prove a lot of bits of $\Omega$, then
$\Omega$ wouldn't be this irreducible
\[
   H(\Omega_N) > N - 8000 .
\]
That would give you a way to compress $\Omega$ into the axioms of the
FAS.  It would give you too concise a way to calculate $\Omega$.  If
you could prove what the bits of $\Omega$ are, you'd do it
systematically by searching through all possible proofs, and that
would give you a way to calculate the bits of $\Omega$.  That's all
we're saying, that it would contradict this
\[
   H(\Omega_N) > N - 8000 .
\]

So deduction and computation are very close, as Turing already noticed
in his famous 1936 paper ``On Computable Numbers...'' where he proves
an incompleteness result using the unsolvability of the halting
problem.  My argument is at the same level, it's analogous.

The new business here is roughly like the paradox of the first
uninteresting positive integer.  You know, that's sort of an
interesting number.  So if you could prove that a number's
uninteresting, that would be an interesting fact about it, and
``uninteresting'' is the notion of algorithmically incompressible.  So
it turns out that you can't prove that an $N$-bit string is
algorithmically incompressible if $N$ is larger than the complexity of
your axioms.  You can't prove that an $N$-bit string is
algorithmically irreducible if it has more bits than the axioms you're
using for the proof.  And similarly it turns out that with $N$ bits of
axioms you get at most $N+15328$ bits of $\Omega$.

That's the general idea.  Here are the details.

I write out a 7328-bit program $\phi_{7328}$, it's about one page of
LISP code.  Why 7328 bits?  Because you have to add that to the
constant in the inequality for $H(\Omega_N)$ to get the constant in
our incompleteness result.  So the difference between these two
constants 8000 and 15328 is the size of the next program in this
course, the last program.  Divided by 8, you get the size of a LISP
expression in characters.

And what this LISP program $\phi_{7328}$ does is this.  Using {\tt
try} it starts running the formal axiomatic system that you're assumed
to be given that's $N$ bits of code.  So we're looking at this:
\[
   U( \phi_{7328} \, \mbox{FAS}^\ast \ldots ) .
\]
The prefix $\phi_{7328}$, the program that outputs the theorems of the
FAS, and some extra stuff that I'll explain later are concatenated and
fed to $U$.

$\phi_{7328}$ starts running the formal axiomatic system using larger
and larger time bounds, capturing the intermediate output, which are
the theorems.  And it looks at the theorems to see how many bits of
$\Omega$ it got.  And I allow partial determinations where you get
some of the bits but you leave holes with unknown bits between some of
the bits that you do know.

And what $\phi_{7328}$ does is it looks for a small set of axioms that
enable you to prove substantially more bits of $\Omega$ than there are
in those axioms---at least 15329 bits more.  And then it just fills in
the holes, the missing bits, which costs just one bit per bit, one bit
for each missing bit.  So $\phi_{7328}$'s final output and the value
of
\[
U( \phi_{7328} \, \mbox{FAS}^\ast \, \mbox{missing bits of $\Omega$} )
\]
will be one of the $\Omega_N$'s.  For some $N$, it'll be the list of
the first $N$ bits of $\Omega$.  By the way, $\phi_{7328}$ may not
need all the bits of FAS$^\ast$, because it only keeps a finite part
of the potentially infinite computation for the formal axiomatic
system.

So if you could use $N$ bits of axioms to get essentially more than
$N$ bits of $\Omega$---more than $N+15328$ bits, in fact---then you
could fill in the missing bits at a cost of one bit each, and you get
into trouble with this inequality
\[
   H(\Omega_N) > N - 8000 .
\]
That's the point.

So that's how it goes, the basic idea is straight-forward.  And I
actually have this 7328-bit LISP program $\phi_{7328}$ and you can run
it.  Well, you can certainly test the auxiliary functions.  You can
certainly run all the auxiliary functions on examples to convince
yourself that they work.  But to test the whole algorithm, its main
function, you have to cheat, you don't give it a real formal axiomatic
system, because if you really try to run through all possible proofs
in size order it would take too long.  So you cheat, you give the
algorithm a LISP expression that just throws out a few test theorems
that will put the algorithm $\phi_{7328}$ that's running the formal
axiomatic system through it's paces.

That's how you convince yourself that this all works.  And you may
have to ``tweak'' things a little bit, you may have to slightly change
the LISP code so that it works with your simple test cases.  So you
can convince yourself by running little examples that this would all
work if you really gave it a genuine formal axiomatic system, say
Zermelo-Fraenkel set theory.

I would like to have somebody program out Zermelo-Fraenkel set theory
in my version of LISP, which is pretty close to a normal LISP as far
as this task is concerned, just to see how many bits of complexity
mathematicians normally assume.  You see, the whole effort here has
been to get these constants 8000 and 15328.  And if you programmed ZF,
you'd get a really sharp incompleteness result.  It wouldn't say that
you can get at most $H(\mbox{ZF})+15328$ bits of $\Omega$, it would
say, perhaps, at most 96000 bits!  We'd have a much more definite
incompleteness theorem.  I hope that somebody will program ZF in LISP\@.
I stop at this point [laughter], you know, programming fatigue!  And
you want to leave something for the students to do, right?  So this is
a little job for them.  And if they're really clever maybe it'll put
out the theorems in Zermelo-Fraenkel set theory reasonably quickly,
not by running through all possible proofs in size order, which would
take too long, but by doing some kind of tree search so that you can
actually have something interesting happen in a reasonable amount of
time.

\section*{Discussion}

Okay, so why is all this interesting?  Well, some of you may know
already, but let me wave my hands frantically in the last few minutes
and say why.

This looks like a kind of programming madness, right?  Up to know, it
looks like what I've been telling you is, ``Oh how much fun it was to
program something that nobody's ever programmed before!''  Or maybe
just that this is neater than it was in the previous versions of my
course.  Yes, programming can be an obsession, it can ruin your life.
You can sit at your terminal watching your life fall apart as you stay
there hacking away until 4 in the morning!  But is there any other
justification for this except as a kind of drug?  Well, I think so!  I
think this has some philosophical significance.

What's the point about this incompleteness theorem, about this
technical result?
\begin{quote}
If a FAS has program-size complexity $N$, then it can enable you to
determine at most $N+15328$ bits of $\Omega$.
\end{quote}

Well, what it's really telling you is that you might get some bits of
$\Omega$, the first ones, say.  You might be able to calculate the
first few bits of $\Omega$.  In fact, in a different version of this
theory, with a different UTM, (this was with my one-character LISP, my
one character per atom LISP), I did get the first 7 bits of $\Omega$,
and they were all 1 bits.  You see, once your lower bound on $\Omega$
is 127/128ths, you know that these bits can't change.  The first 7
bits were 1's.  And now I have a much friendlier LISP, but I lost
this, I can no longer determine the first 7 bits of $\Omega$ this way.

But anyway, you might be able to get some bits of $\Omega$ without
contradicting $\Omega$'s logical and computational irreducibility,
without contradicting this incompleteness result and this inequality:
\begin{itemize}
\item
You can determine at most $H(\mbox{FAS})+15328$ bits of $\Omega$.
\item
$H(\Omega_N) > N - 8000.$
\end{itemize}

But in spite of this, the basic point is that $\Omega$ really shows
that some areas of mathematics have no structure, have no pattern at
all.

Let me put it this way.  Normally you think that if something is true,
it's true for a reason, right?  In mathematics, the reason is called a
proof, and the job of a mathematician is to find the reason that
something is true, to find a proof.  But the bits of $\Omega$ are
mathematical facts that are true for no reason, they're accidental!

Now this is a very specific $\Omega$, I've programmed it out.  Imagine
it being written in binary.  And you ask, an individual bit, say the
33rd bit, is it a 0 or a 1?  Let's say you're trying to prove which
it is.  And the answer is, you can't!  The reason you can't is
because, whether that particular bit is a 0 or a 1 is true for no
reason, it's true by accident.  It's so delicately balanced whether
it's going to be a 0 or a 1, that we will never know!

It's like independent tosses of a fair coin.  Independent tosses of a
fair coin has got to come out heads or tails in each case, but there's
no reason that it comes out one or the other, right?  So it's exactly
the same story with these mathematical facts, with the bits of
$\Omega$.  There is no pattern or structure in the sequence of bits of
$\Omega$.

I don't know why anybody would want to try to determine bits of
$\Omega$.  Although people have played with the Busy Beaver function,
which in a way is like trying to calculate the bits of $\Omega$.  I
don't know why you'd try to determine the bits of $\Omega$.  But if
you were to try to do this, what this incompleteness result shows you
is that you're in big, big trouble!  Because essentially the only way
to prove what a bit of $\Omega$ is, is to add the theorem that you
want to prove as a new axiom.  It's irreducible mathematical
information.  Now you can prove {\bf anything} by adding it as a new
axiom.  The point here is that for $\Omega$ that's essentially {\bf
the only way} to do it.  No compression is possible.

So there is no structure, there is no pattern, $\Omega$ has maximum
entropy, it mirrors independent tosses of a fair coin.  Now to a
physicist what I'm saying sounds pretty reasonable, right?  $\Omega$
has maximum entropy, the bits of $\Omega$ are completely uncorrelated.
But to a mathematician this all sounds weird!

My new $\Omega$ is a particular well-defined real number.  It's a real
number with a rather simple definition, I can even write the LISP
program that defines $\Omega$ on one computer screen.  So you believe,
thinking in Platonic terms, that each bit is either a 0 or a 1, even
if I'm never going to know which.  But it's black or white, right?
And what I'm saying is that I think that it's really better to think
that it's {\bf grey}.  It's really better to think that each bit of
$\Omega$ has probability one-half of being 0 or of being 1---even
though it's a particular well-determined bit, because I've written out
the program that defines this number.  Defines it, not by enabling you
to calculate it bit by bit---that would contradict $\Omega$'s
unknowability.  But the program for $\Omega$ does enable you to
calculate better and better lower bounds on $\Omega$, so $\Omega$ is
{\bf almost} a computable real number in the same sense that $\pi$ is.
Almost, but not quite!

So the game is like this.  I'm trying very hard to be constructive, as
constructive as possible.  Writing out programs is a sure sign that
you're a constructivist, right?  [Laughter] I want to settle all the
programming details.  But I'm trying to be as constructive as possible
about non-constructivity!  I want to exhibit something that escapes
the power of constructivity, of mathematical reasoning, that you can't
calculate, but that's just over the border between the constructible
and the non-constructible.  And I think that $\Omega$ is pretty damn
good, it's just on the border.

Now this doesn't mean that all of mathematics falls down in a heap!
But the normal notion of mathematics was that there were a small,
finite set of axioms and rules of inference that we could all agree
on, from which all the infinite mathematical truth would follow.  This
is the tradition that goes back to Euclid, to Leibniz, to Peano, to
Frege, to Russell and Whitehead, to Hilbert.  And G\"odel showed that
there was a problem.  And Turing showed that there was a problem,
using a different method involving computers.  And I think that
$\Omega$ follows in that tradition and shows that the problem is even
bigger.  However I don't think this means that you should stop doing
mathematics!  So what is the significance of $\Omega$ and of
incompleteness?  Should it affect how we actually do mathematics?
I'll give you my opinion.

Of course the nature of mathematics has been discussed for a long
time!  Every generation of mathematicians have their own answer.  But
let me share with you my own feelings about this, my tentative
conclusions.

There's a word that a philosopher coined that's very good.  He says
that there's an emerging new school, and emerging new {\it
quasi-empirical\/} view of the foundations of mathematics.  One talks
about the formalist school, the logicist school, and the intuitionist
school.  Well, Thomas Tymoczko has a book, {\it New Directions in the
Philosophy of Mathematics,} which has a whole bunch of articles,
including two of mine, and he thinks that all these articles tend to
support a new quasi-empirical view of the foundations of mathematics.

Now what does quasi-empirical mean?  I'll tell you what it means to
me.  Quasi-empirical means that {\bf pure math ain't that different
from physics!} The normal notion of pure math is that mathematicians
have some kind of direct pipeline to God's thoughts, to absolute
truth!  But poor physicists!  You know, they tried Newtonian
mechanics, it looks good for a while, then Einstein shows it's all
wrong.  Then---surprise!---quantum mechanics shows that Einstein was
wrong!  And now there's superstring theory.  And is it right?  Is it
wrong?  And mathematicians laugh and say, ``Oh those poor physicists!
It's such a messy subject!  They always have to backpedal, they don't
know what they're doing!  It's all so tentative!''

Well, I think that mathematics and physics are not really that
different!---Physicists love it when I say this!  [Laughter]

Let me try explaining this another way.  Euclid said that mathematics
is based on self-evident truths.  But my impression is that maybe
axioms are not self-evident truths.  I don't believe in self-evident
truths.  Maybe it's more like in physics.  Maybe mathematics should be
done more like physics, where you're willing to add new axioms because
they're useful, not because they're self-evident.  And then of course
you have to be prepared to say ``I goofed!''\ and remove an axiom,
which mathematicians don't like to have to do.

This may sound completely crazy, but in fact it's not just my opinion.
G\"odel makes very similar remarks in Volume II of his {\it Collected
Works,} in his essay on ``Russell's Mathematical Logic.''  This essay
was originally published in the Paul Arthur Schilpp volume {\it The
Philosophy of Bertrand Russell.}

I talked about new axioms to a mathematician once, and he replied,
``Okay, I'm willing to add the Riemann hypothesis as a new axiom if
you can prove to me that it doesn't follow from the usual axioms.''
Well, that's hard to do, because if the Riemann hypothesis were false,
then there would be a numerical counter-example that one could easily
verify that shows that it's false.  So if you could show that the
Riemann hypothesis is beyond the power of the usual axioms, that would
imply that the Riemann hypothesis is true!

So these ideas are very controversial.

But the whole point of algorithmic information theory, the whole point
of my information-theoretic approach to incompleteness, is that
sometimes to get more information out of a set of axioms, you've just
got to put more in.  So let's put more axioms in!  Physicists have
always done that.

I had these ideas a long time ago.  I proved my first
information-theoretic incompleteness theorem in 1970.  Although it's
only in the past two or three years that I discovered how to actually
program the algorithm in my original proof and run it on examples.
And I was recently surprised to discover or rediscover that there are
highly relevant quotes from Einstein and from G\"odel.  Let me throw
them into this discussion, let me end with that.

Einstein has a very nice remark that I angered some mathematicians
with.  But what do they care, he's only a physicist, right?
[Laughter] In his essay ``Remarks on Bertrand Russell's Theory of
Knowledge'' in the Paul Arthur Schilpp volume {\it The Philosophy of
Bertrand Russell,} Einstein says that ``the series of integers is
obviously an invention of the human mind, a self-created tool which
simplifies the ordering of certain sensory experiences.''  So you can
see that Einstein's attitude is very empirical.

I think that Einstein's position is that the positive integers are not
{\it a priori,} they're not God given, we invent them like we invent
all of physics.  But the positive integers look more {\it a priori\/}
than other concepts because they've been around longer, we invented
them a long time ago.  After all, when an idea has been around for a
few thousand years, it's not surprising that people think that it's
obvious, that it's just common sense, that it's ``a necessary tool of
thought.''  The other extreme is the latest field theory in physics.
That looks a lot more tentative.  It hasn't been here long, and it'll
probably be shot down next week, right?  And there are probably
thirteen different versions!  But in Einstein's view there is no
fundamental difference.  The positive integers have been around for a
long time, but they're still just an invention.

So I like this quote from Einstein, but it doesn't convince
mathematicians.

Then there are some very interesting remarks made by G\"odel.  You can
find them in G\"odel's {\it Collected Works.} G\"odel's philosophical
position is the exact opposite of Einstein's.  G\"odel believed in the
Platonic universe of mathematical ideas, he believed that the positive
integers are just as real as tables and chairs.  There are an infinity
of positive integers, and they're out there somewhere.  They're in the
Platonic universe of mathematical ideas, that's where mathematical
objects are!

I don't know!  I used to laugh at all of this when I was a kid.  But
think about it seriously.  You're young, you're trying to learn
mathematics, you're doing elementary number theory, and you don't
really start to worry about the fact that elementary number theory
presupposes arbitrarily large positive integers, and how do they fit
in the universe?  Imagine a positive integer that is
\[
10^{10^{10^{10}}}
\]
digits long.  Does it exist?  In what sense does it exist?  You don't
care---right?---you prove theorems about it, you know that it would be
commutative, right?  $a+b$ is going to be equal to $b+a$ even if
neither number fits in the universe!  [Laughter] But then later on in
life you start to worry about this!  [Laughter]

So I'm not sure if the positive integers exist anymore.  But G\"odel
thinks that they do, and that philosophical position is associated
with the name of Plato.  But it's really just the classical
mathematical position that the positive integers really exist, that
an infinity of them are really out there.  And starting from that,
G\"odel comes to a very surprising conclusion.  Since the positive
integers are just as real as tables and chairs, you can do experiments
with them by doing calculations, and if you see a pattern, you can
just go ahead like a scientist would in dealing with electrons.  Since
integers are just as real as electrons, why can't we use the same
kinds of methods that scientists use?

Here are G\"odel's exact words, taken from a previously unpublished
manuscript {\it *1951\/} that is in Volume III of his {\it Collected
Works:}
\begin {quote}
   ``If mathematics describes an objective world just like
   physics, there is no reason why inductive methods should
   not be applied in mathematics just the same as in physics.''
\end{quote}

And in his essay ``What is Cantor's Continuum Problem?''\ in Volume II
of his {\it Collected Works,} G\"odel says that maybe we'll come up
with new versions of set theory, maybe we'll come up with new ideas
about sets the same way that physicists come up with new ideas about
physical objects.  The justification for these new principles would be
that they're useful, that they help us to organize our mathematical
experience, just as the ultimate justification for physical principles
is that they help us to organize our physical experience.

I think it's very funny!  Here's Einstein, who's a diehard empiricist,
and here's G\"odel, who's a diehard Platonist, and they sort of come
to the same conclusion!  But they were buddies at the Institute for
Advanced Study, so it's not surprising that they influenced each
other.  And the information-theoretic viewpoint that I've explained to
you today leads me in the very same direction.

Another funny thing is that I don't believe that all this work on the
foundations of mathematics has changed at all the way mathematicians
actually work.  G\"odel's incompleteness theorem was initially very
shocking.  But then mathematicians noticed that the kind of assertion
that G\"odel constructed that's true but unprovable is not the kind of
assertion that you deal with in your everyday work as a mathematician.
And I think it's fair to say that about $\Omega$ too.  The longer I
live with $\Omega$, the more natural it looks to me.  But a skeptic
would say, ``I don't care about the bits of $\Omega$, so what if
there's trouble there!''

But I do think that something else is making a difference in how
mathematicians carry on their everyday work.  It's the fact that the
computer has so vastly expanded mathematical experience, that we just
have to deal with it somehow.  A lot of physicists do experimental
work on the computer now.  There's even a journal called {\it
Experimental Mathematics.}

The computer has so vastly increased mathematical experience, that how
do we keep up with all of it?  Well, the answer is that sometimes you
see that something seems to be the case, and it'd be nice if you could
prove it, but for the moment you can't, so you sort of conjecture it.
And if you're doing mathematical physics, then you're on the
borderline between math and physics, and it's certainly okay to behave
in this way.  But if you're a mathematician, if you're just over the
border, then it's not so clear anymore.

Well, I once had a conversation like this with a mathematician, and
the Riemann hypothesis came up, and he said, ``It works fine the way
we do things now.  You have a paper, and you say the paper is `modulo
the Riemann hypothesis.'  So why do you need to call it a new axiom?''
And he has a point.  But I do think that if a principle is very
helpful for a long time, and nobody shoots it down, why not call it a
new axiom?  But one shouldn't do this too quickly.  One has to be
careful, the way physicists are, although I sometimes wonder how
careful physicists really are!  [Laughter]

Okay, that's the idea.  And if you go to my Web site you'll find there
in HTML all my less technical papers.  And you've also got this course
there, and the Java source code and byte code for my LISP interpreter.
So you're welcome to go to my Web site and play with it, and you're
welcome to send me e-mail about it, and I'd be pleased as Punch if one
of you tried giving my course on the limits of mathematics to real
students in a normal university setting.

Anyway, you guys have been nice enough to invite me for many years to
give talks, and it's been very stimulating to talk with you, and for
the moment I think I've exhausted this train of thought.  Of course,
I've been thinking that since I was fifteen years old, but fortunately
every now and then I get a new idea!  [Laughter] But for now, that's
all I have to say, and maybe someday I'll have more to say, and I'll
be lucky enough to come out here again and have a chance to kick it
around with you!  Thank you very much!  [Applause]

\end{document}